\documentclass[12pt]{iopart}
\usepackage{epsfig}
\usepackage{graphicx}
\usepackage{iopams}
\usepackage{wrapfig}
\begin{document}
\title[Single Freeze-Out, Statistics and Identified Hadron Production]
{Single Freeze-Out, Statistics and Pion, Kaon and Proton Production in
Central Pb-Pb Collisions at $\sqrt{s_{NN}} = 2.76$ TeV}

\author{Dariusz Prorok}

\address{Institute for Theoretical Physics, University of
Wroc{\l}aw, Pl.Maksa Borna 9, 50-204 Wroc{\l}aw, Poland}
\ead{dariusz.prorok@ift.uni.wroc.pl}
\vspace{10pt}
\begin{indented}
\item[]March 2016
\end{indented}
\begin{abstract}
Many data in the high energy physics are, in fact, sample means. It is shown that when this exact meaning of the data is taken into account and the most weakly bound states are removed from the hadron resonance gas, the acceptable fit to the whole spectra of pions, kaons and protons measured at midrapidity in central Pb-Pb collisions at $\sqrt{s_{NN}} = 2.76$ TeV [Phys.Rev.Lett.{\bf 109},252301(2012)] is obtained. The invariant distributions are predicted with the help of the single-freeze-out model in the chemical equilibrium framework. Low $p_{T}$ pions and protons are reproduced simultaneously as well as $p/\pi$ ratio. Additionally, correct predictions extend over lower parts of large $p_{T}$ data. Some more general, possible implications of this approach are pointed out.
\end{abstract}

%
\vspace{2pc}

%
%
%

\section{Introduction}
\label{intro}

High-energy heavy-ion collisions are the tools for the creation of the deconfined phase (the partonic system) of the Quantum Chromodynamics (QCD) (for a wide review of the subject, from the theory to the experiment, see Ref.~\cite{Leupold:2011zz}). The matter originated during such a collision, extremely dense and hot, is compressed more or less in the volume of the narrow disc of the ion radius at the initial moment. After then the matter rapidly expands due to the tremendous pressure and cools simultaneously. The evolution of the matter can be described in the framework of the relativistic hydrodynamics \cite{Huovinen:2013wma}. During expansion the matter undergoes a transition to a hadron gas phase. The hadron gas continues the hydrodynamical evolution, assuming that the collective behavior does not cease at the transition. The expansion makes the gas more and more diluted, so when mean-free paths of its constituents become comparable to the size of the system one can not treat the gas as a collective system. This moment is called \textit{freeze-out}. After then the gas disintegrates into freely streaming particles which can be detected. In principle, one can distinguish two kinds of freeze-out: \textit{a chemical freeze-out}, when all inelastic interactions disappear  and \textit{a kinetic freeze-out} (at lower temperature), when also elastic interactions disappear. The measured hadron yields are fingerprints of corresponding hadron abundances present at the chemical freeze-out. The yields can be consistently described within the grand canonical ensemble with only two independent parameters, the chemical freeze-out temperature, $T_{ch}$ and the baryochemical potential $\mu_B$ \cite{BraunMunzinger:2003zd}. This idea is the fundament of the Statistical Model (SM) of particle production in heavy-ion collisions. The measured $p_T$ spectra include information about the transverse expansion (radial flow) of the hadron gas and the temperature $T_{kin}$ at the kinetic freeze-out \cite{Heinz:2004qz}. However, the alternative approach to freeze-out was founded in \cite{Broniowski:2001we,Broniowski:2001uk} where the single freeze-out was postulated, i.e. the kinetic freeze-out coincided with the chemical freeze-out. This is the Single-Freeze-Out Model (SFOM). The suitably chosen freeze-out hypersurface and the complete inclusion of contributions from resonance decays enabled to correctly describe the Relativistic Heavy Ion Collider (RHIC) $p_T$ spectra.

With the first data on Pb-Pb collisions at $\sqrt{s_{NN}} = 2.76$ TeV from CERN Large Hadron Collider (LHC) \cite{Abelev:2012wca,Abelev:2013vea} two new problems have appeared when the SM and hydrodynamics were applied for the description of particle production. The predicted proton and antiproton abundances were much larger then measured ones \cite{Stachel:2013zma} and low $p_{T}$ pions were underestimated \cite{Abelev:2012wca,Begun:2013nga,Melo:2015wpa}. This caused that the ratio $p/\pi = (p + \bar{p})/(\pi^{+} + \pi^{-})$ was overestimated in the SM by a factor $\sim 1.5$ \cite{Floris:2014pta}. Various explanation of this ''puzzle'' have been invented, but all fall outside the SM. These are: (i) the incomplete list of resonances, there could still be undiscovered (high mass) resonances which after decays would increase more pion yields than proton ones, (ii) the non-equilibrium thermal model, with two additional parameters describing the degree of deviation from the equilibrium, (iii) hadronic inelastic interaction after hadronization and before chemical freeze-out, especially baryon annihilation, and (iv) flavor hierarchy at freeze-out, which could result in two different freeze-out temperatures, one for non-strange hadrons, another for strange hadrons (for more details and references see \cite{Floris:2014pta}).

In this work the simple generalization of the SFOM in the chemical equilibrium framework is postulated, where the above  problems disappear naturally and whole results (spectra and yields) of \cite{Abelev:2012wca} are reproduced. However, in opposite to the original version, all parameters of the model (thermal and geometric) are estimated simultaneously from the spectra. This version was successfully applied to the description of the final spectra measured at RHIC for all centrality classes in the broad range of collision energy \cite{Prorok:2007xp}. The new idea introduced into the SFOM in the present work is to randomize one of the parameters of the model. It has turned out that the successful improvement is achieved only when the freeze-out temperature becomes a random variable and nothing is gained with the randomization of geometric parameters of the model. This result suggests that the temperature of the thermal system at the freeze-out fluctuates significantly in the most central bin of Pb-Pb collisions at $\sqrt{s_{NN}} = 2.76$ TeV but its size remains the same.

\section{The Model}
\label{model}

In the SFOM the invariant distribution of the measured particles of species $i$ has the form

\begin{equation}
{ \frac{dN_{i}}{d^{2}p_{T}\;dy} }=\int
p^{\mu}d\sigma_{\mu}\;f_{i}(p \cdot u) \;, \label{Cooper}
\end{equation}

\noindent where $d\sigma_{\mu}$ is the normal vector on a
freeze-out hypersurface, $u_{\mu}=x_{\mu}/\tau_f$
is the four-velocity of a fluid element at the freeze-out and $f_{i}$ is the final
momentum distribution of the particle in question. The final
distribution means that $f_{i}$ is the sum of primordial and
decay contributions to the distribution. The freeze-out hypersurface is defined by the equations

\begin{equation}
\tau_f = \sqrt{t^2-x^2-y^2-z^2}\;,\;\;\;\sqrt{x^2+y^2} \leq \rho_{max} \;,
\label{Hypsur}
\end{equation}

\noindent where the invariant time, $\tau_f$, and the transverse size, $\rho_{max}$, are two geometric parameters of the model. For the LHC energies all chemical potentials can be put equal to zero, so the freeze-out temperature, $T_f$, is the only thermal parameter of the model. The contribution from the weak decays concerns (anti-)protons mostly \cite{Abelev:2013vea,Milano:2012eea}, hence secondary (anti-)protons from primordial and decay $\Lambda$($\bar{\Lambda}$)'s are subtracted. Fitting expression (\ref{Cooper}) to the all spectra reported in \cite{Abelev:2012wca} (and within \textit{whole ranges} presented there) resulted in $\chi^{2}/n_{dof}$=1.74 with $\emph{p-value} = 2 \cdot 10^{-11}$ ($n_{dof}=235$), which is unacceptable.

However, the data on $p_{T}$ spectra \cite{Abelev:2012wca,Abelev:2013vea} are not 'points' but averages over all events in a sample \footnote{the sample is the 0-5\% centrality class here.}, the division by $N_{ev}$, the number of events in the sample, means that. But the model prediction, Eq.~(\ref{Cooper}), represents a quantity obtained in one collision (one event). Therefore, the theoretical prediction should be also an average. For that reason it is postulated that the expression given by Eq.~(\ref{Cooper}) becomes a statistic (a function of a random variable, by definition it is also a random variable) and that one of the parameters of the model, $\theta$ ($\theta = T_f, \tau_f \; \textrm{or}\; \rho_{max}$), is a random variable. Then the theoretical prediction is defined as the appropriate average:
\begin{equation}
\left\langle { \frac{dN_{i}}{d^{2}p_{T}\;dy} } \right\rangle_{\theta} = \int { \frac{dN_{i}}{d^{2}p_{T}\;dy} } f(\theta) d\theta \;, \label{AvCoop}
\end{equation}
\noindent where $f(\theta)$ is the probability density function (p.d.f.) of $\theta$. This approach is more general but includes the standard one, if fluctuations of $\theta$ are negligible, then its p.d.f. is Dirac-delta like, $f(\theta) \sim \delta(\theta-\theta_o)$ and the average becomes the value at the optimal point $\theta_o$. It has turned out that only randomization of $T_f$ improves the quality of the fit, randomization of $\rho_{max}$ or $\tau_f$ does not change anything. In fact, for the technical reasons, not $T_f$ is randomized but $\beta_f = 1/T_f$. From the statistical point of view these two possibilities are equivalent, because $\beta_f(T_f)$ has a unique inverse and vice verse \cite{Cowan:1998ji}. Two p.d.f.'s are considered: log-normal
\begin{equation}
f(\beta_f;\mu,\sigma) = \frac{1}{\sqrt{2\pi}\sigma}\frac{1}{\beta_f}
\exp \left\{-\frac{(\ln{\beta_f}-\mu)^2}{2\sigma^2} \right\} \; \label{lognorm}
\end{equation}
\noindent and triangular
\begin{equation}
f(\beta_f;\breve{\beta}_f,\Gamma) = \cases{ \frac{\Gamma - \mid \beta_f-\breve{\beta}_f \mid}{\Gamma^2}&for $\mid \beta_f-\breve{\beta}_f \mid \leq \Gamma$  \\  0&for $\mid \beta_f-\breve{\beta}_f \mid > \Gamma \;$.\\ }  \label{triang}
\end{equation}
\noindent where $\mu$ and $\sigma$ are parameters of the log-normal p.d.f. whereas $\breve{\beta}_f$ and $\Gamma$ are parameters of the triangular p.d.f., $\breve{\beta}_f$ is the average of $\beta_f$. The first is differentiable but has an infinite tail, the second is not differentiable but has a finite range. The choice is arbitrary, but two general conditions should be fulfilled, a p.d.f. is defined for a positive real variable and has two parameters so as the average and the variance can be determined independently.

However, in both cases of p.d.f.'s, Eqs.~(\ref{lognorm}) and (\ref{triang}), fits of expression (\ref{AvCoop}) to the whole data on $p_{T}$ spectra \cite{Abelev:2012wca} resulted in $\chi^{2}/n_{dof} = 1.49$ with $\emph{p-value} = 2 \cdot  10^{-6}$ ($n_{dof}=234$), which is still unacceptable.

The second assumption of the model is purely \textit{heuristic} - it states that the most weakly bound resonances should be removed from the hadron gas. To be more precise, all resonances with the full width $\Gamma > 250$ MeV (and masses below 1600 MeV) are removed \cite{Agashe:2014kda}. These are: $f_0(500)$, $h_1(1170)$, $a_1(1260)$, $\pi(1300)$, $f_0(1370)$, $\pi_1(1400)$, $a_0(1450)$, $\rho(1450)$, $K^*_0(1430)$ and $N(1440)$ \footnote{In fact, the hint for this assumption was the accidental observation that after up-to-date of the $f_0(500)$ mass to the lower one \cite{Agashe:2014kda}, the quality of fit became worse. \label{przyp1}}. It should be noticed that the note attached to $f_0(500)$ says: ''The interpretation of this entry as a particle is controversial'' \cite{Agashe:2014kda} and the removal of this resonance has found the theoretical justification recently \cite{Broniowski:2015oha}. The exclusion of only $f_0(500)$
moves fits to the boundary of the acceptance, $\chi^{2}/n_{dof} \sim 1.3$ ( $\emph{p-value} \sim 0.001$), nevertheless according to the rigorous rules of the statistical inference it is still not a ''good'' fit \cite{Cowan:1998ji}. The removed resonances are weakly bound already in the vacuum, with the average lifetime $\tau < 1$ fm, so they might not form in the hot and dense medium at all, at least in the case of central Pb-Pb collisions at extreme energy $\sqrt{s_{NN}} = 2.76$ TeV. Anyway, this is a \textit{heuristic} hypothesis, but it works very well. It should be stressed at this point that both assumptions are necessary, if only the removal of weakly bound resonances is applied (no randomization of any parameter), the fit is still unacceptable, $\chi^{2}/n_{dof} = 1.5$ ($\emph{p-value} =  10^{-6}$). It looks like both assumptions (phenomena) strengthen each other.

\section{Results}
\label{result}

\begin{figure}[h]
\includegraphics[width=0.7\textwidth]{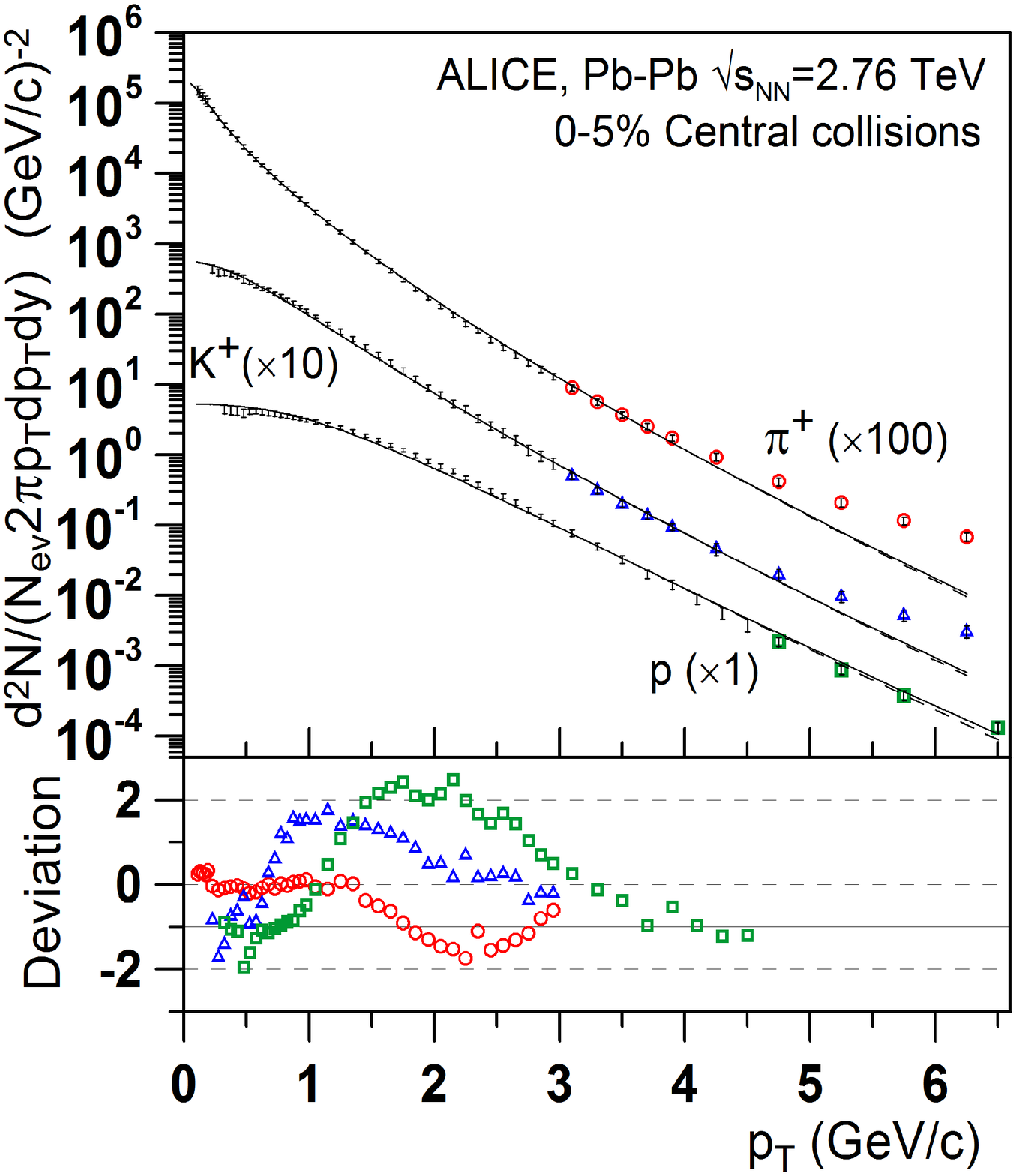}
\caption{\label{Fig.1} The upper panel presents spectra of positive pions, kaons and protons measured in central Pb-Pb collisions at $\sqrt{s_{NN}} = 2.76$ TeV, data used in the fit are presented as error bars only \protect\cite{Abelev:2012wca}, errors are sums of statistical and systematic components added in quadrature. Symbols denote one half of corresponding sums of negative and positive hadrons from large $p_T$ measurements \protect\cite{Abelev:2014laa}. Lines are predictions for the log-normal p.d.f. of $\beta_f$, dashed lines for the triangular p.d.f., in the fitted region they cover each other. The lower panel shows a deviation of data to the model in the fitted range measured in error units, $(f_{exp}-f_{mod})/\sigma_{exp}$, where $f_{exp(mod)}$ is the experimental (model) value of the invariant yield at given
$p_{T}$ and $\sigma_{exp}$ is the error on $f_{exp}$. In both plots circles (red online) denote pions, triangles (blue online) kaons and squares (green online) protons.}
\end{figure}

\begin{figure}[h]
\includegraphics[width=0.7\textwidth]{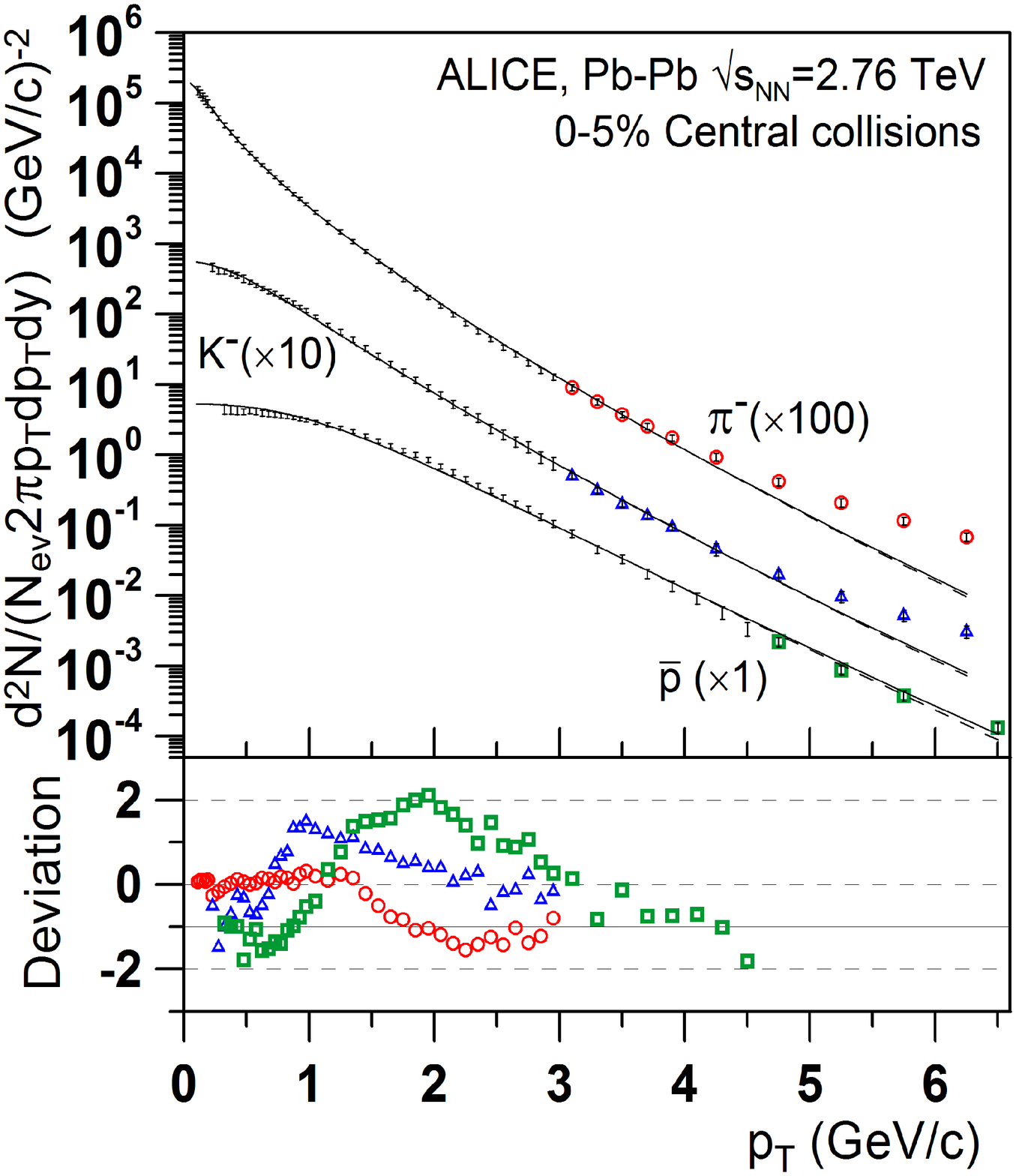}
\caption{\label{Fig.2} The same as Fig.\,\ref{Fig.1} but for negative pions, kaons and antiprotons. }
\end{figure}

The results of fits are presented in Table~\ref{Table1} and depicted in Figs.\,\ref{Fig.1} and \ref{Fig.2}. Predicted spectra are the same for positive charge particles and corresponding negative charge particles. This is because, when all chemical potentials equal zero, particles are distinguished from each other only by a mass. But particles and their antiparticles have the same mass and additionally patterns of decays of antiparticles are mirrors of patterns of decays of corresponding particles - the hadron resonance table consists of particles together with their antiparticles.

\begin{table}[hb]
\caption{\label{Table1} Fit results for 0-5\% central Pb-Pb collisions at $\sqrt{s_{NN}} = 2.76$ TeV and the measurement at central rapidity, $\mid y \mid < 0.5$, $n_{dof}=234$. Parameters of the log-normal p.d.f. have no units, so their values correspond to $\beta_f$ counted in MeV$^{-1}$ implicitly. }
\footnotesize
\begin{tabular}{@{}llllllll}
\br
 \multicolumn{8}{c}{log-normal p.d.f.}
\\
 \cline{1-8}
 $\tau_f$ & $\rho_{max}$ & $\mu$ & $\sigma$ & $E[T_f]$ & $\sqrt{V[T_f]}$ & $\chi^{2}/n_{dof}$ & $\emph{p-value}$
\\
 $(\textrm{fm})$ & $(\textrm{fm})$ &  &  & (MeV) & (MeV) &  & (\%)
\\
\mr 13.80 $\pm$ 0.40 & 20.48 $\pm$ 0.60 & -4.7439 $\pm$ 0.0235 & 0.1764 $\pm$ 0.0090 & 116.7 $\pm$ 3.0 & 20.7 $\pm$ 1.6 & 1.048 & 29
\\
\br
 \multicolumn{8}{c}{triangular p.d.f.}
\\
 \cline{1-8}
 $\tau_f$ & $\rho_{max}$ & $\breve{\beta}_f$ & $\Gamma$ & $E[T_f]$ & $\sqrt{V[T_f]}$ & $\chi^{2}/n_{dof}$ & $\emph{p-value}$
\\
 $(\textrm{fm})$ & $(\textrm{fm})$ & (MeV$^{-1}$) & (MeV$^{-1}$) & (MeV) & (MeV) & & (\%)
\\
\mr 14.42 & 21.45 & 0.0092482 & 0.0040906 & 111.6 & 22.6 & 1.026 & 38
\\
\br
\end{tabular}
\end{table}


Note that errors on estimates of parameters are given only for the log-normal p.d.f.. This is because the standard way of expressing these errors is via the Hessian of the $\chi^{2}$ test statistic evaluated at the estimates \cite{Cowan:1998ji}. Precisely, the inverse covariance matrix is given by one half of the Hessian. This requires the $\chi^{2}$ statistic to be at least twice differentiable with respect to parameters, which is not fulfilled in the case of the triangular p.d.f.. Both fits are acceptable and with the same quality practically. Predictions of both cover each other in the whole fitted range and even further, they start to deviate from each other at $p_T \approx 4.5$ GeV/c, see Figs.\,\ref{Fig.1} and \ref{Fig.2}. As it can be seen from the bottom plots, predicted values stay within the two error band in general, only a few protons in the range 1.5-2.5 GeV/c exceed the band slightly. Additionally, the predictions extend over lower part of the large $p_T$ measurements \cite{Abelev:2014laa}, pions are reproduced up to $p_T \approx 4$ GeV/c, kaons up to $p_T \approx 4.5$ GeV/c and protons and antiprotons up to $p_T \approx 6$ GeV/c. In Figs.\,\ref{Fig.1} and \ref{Fig.2} data from the large $p_T$ measurements are presented as one half of corresponding sums of negative and positive hadrons reported in Ref.~\cite{Abelev:2014laa}.

In Table~\ref{Table1} the expectation value, $E[T_f]$, and the square root of the variance, $\sqrt{V[T_f]}$, of the freeze-out temperature are given. The expectation value is of the order of 115 MeV, whereas the square root of the variance is $\sim 20$ MeV. The latter is the measure of the fluctuations of the freeze-out temperature in the sample and is of the order of 20\%. Since the distribution of the freeze-out temperature means the distribution within the sample in this work, these fluctuations reflect the changes of the freeze-out temperature from one event to another. Generally, these changes could be of the \textit{thermal} and \textit{non-thermal} origin. The thermal component represents fluctuations in an ensemble as derived e.g. in \cite{Landau:1980}, i.e. it describes changes of the temperature from one system to another but prepared exactly in the same way. The non-thermal component expresses the variation of the freeze-out conditions event-by-event. The occurrence of this variation would mean that systems (events) were prepared differently in spite of the fact that they belong to the same centrality class. This would indicate that the 0-5\% centrality class is inhomogeneous significantly. With the data available at present, it is impossible to distinguish between these two components. The inhomogeneity might be expected from the analysis of the centrality determination \cite{Abelev:2013qoq}, where the estimates of ranges of the impact parameter for centrality classes are given. In the 0-5\% centrality class this parameter varies from 0 to 3.5 fm and this is the biggest spread for all classes considered in \cite{Abelev:2013vea}, for the rest the spread is 2 fm at most \cite{Abelev:2013qoq}. However, because the impact parameter determines the initial geometry of the collision, one could expect that its variation would influence the final geometry of the collision, i.e. $\rho_{max}$ should vary from one event to another, rather. Results of this work show that this is not the case, randomization of $\rho_{max}$ (or $\tau_f$) has not improved the quality of fits, they remained unacceptable. Another possibility is that the variations of the final size of the system within the sample have negligible impact on the spectra. Results of this model suggests that during each event a thermal system is created indeed and with approximately the same size at its end, however with different temperature. And the final shape of the spectra is the consequence of summing emissions from many different sources.

In the SFOM particle yields per unit rapidity are given by \cite{Begun:2014rsa}:
\begin{equation}
{ \frac{dN_{i}}{dy} }= \pi \rho_{max}^2 \tau_f n_i\;, \label{Yields}
\end{equation}
\noindent where $n_i$ is the thermal density of particle species $i$. The above expression can be obtained by the integration of the distribution (\ref{Cooper}) over transverse momentum. But here predictions are appropriate averages, so the average particle yield per unit rapidity reads:
\begin{equation}
\left\langle{ \frac{dN_i}{dy} }\right\rangle_{\beta_f} = \pi \rho_{max}^2 \tau_f \left\langle n_i\right\rangle_{\beta_f} \;, \label{AveYiel}
\end{equation}
\noindent because integration over $p_T$ can be exchanged with the integration over $\beta_f$ and $\rho_{max}$ and $\tau_f$ are independent variables. Results for the yields calculated with the use of Eq.~(\ref{AveYiel}), at the values of parameters gathered in Table~\ref{Table1}, are given in Table~\ref{Table2}. Excellent agreement with the data \cite{Abelev:2012wca} has been achieved. Since yields are correctly reproduced, their ratios are correct as well. The conclusion of Ref.~\cite{Broniowski:2015oha} about the enhancement of $p/\pi$ ratio after removal of $f_0(500)$ state should be commented at this point. This is true, but at the same values of model parameters as before the removal, i.e. at the constant temperature or at the constant parameters of the temperature distribution. However after the removal, the expression for the model prediction, Eqs.~(\ref{Cooper}) and (\ref{AvCoop}), changes to the new one, so the $\chi^{2}$ test statistic changes as well and the new minimization is necessary. This new minimization estimates new values of parameters and this analysis proves that the new $p/\pi$ ratio is smaller.

\begin{table}[hb]
\caption{\label{Table2} Midrapidity particle yields $\frac{dN_i}{dy}|_{|y|<0.5}$ and their ratios. Data are from \protect\cite{Abelev:2012wca}.}
\begin{indented}
\item[]\begin{tabular}{@{}llll}
\br
 & & \multicolumn{2}{c}{Model: $\left\langle{ \frac{dN_i}{dy} }\right\rangle_{\beta_f}$}
\\
 \cline{3-4} Species & Data & triangular p.d.f. & log-normal p.d.f.
\\
\mr
 $\pi^{+}$ & 733 $\pm$ 54.0 & 745.3 & 739.2
\\
$\pi^{-}$ & 732 $\pm$ 52.0 & 745.3 & 739.2
\\
$K^{+}$ & 109 $\pm$ 9.0 & 106.9 & 107.5
\\
$K^{-}$ & 109 $\pm$ 9.0 & 106.9 & 107.5
\\
p & 34 $\pm$ 3.0 & 33.0 & 32.9
\\
$\bar{p}$ & 33 $\pm$ 3.0 & 33.0 & 32.9
\\
\br
 \multicolumn{4}{c}{Ratios}
\\
 \mr
    & Data & triangular p.d.f. & log-normal p.d.f.
\\
\mr
 $p/\pi$ & 0.046 $\pm$ 0.003 & 0.044 & 0.045
\\
 $K/\pi$ & 0.149 $\pm$ 0.010 & 0.143 & 0.145
\\
\br
\end{tabular}
\end{indented}
\end{table}

\section{Conclusions}
\label{Conclus}

In summary, the chemical equilibrium Single-Freeze-Out Model has been applied successfully to the description of the production of identified hadrons measured at midrapidity in central Pb-Pb collisions at $\sqrt{s_{NN}} = 2.76$ TeV \cite{Abelev:2012wca}. This has been achieved with the help of the more general, direct interpretation of the data and the removal of the most weakly bound resonances from the hadron gas. Since the chemical equilibrium SFOM without the above two new assumptions succeeded in the correct description of spectra measured at the RHIC in the broad range of collision energy \cite{Broniowski:2001we,Broniowski:2001uk,Prorok:2007xp,Prorok:2005uv}, it might suggest new phenomena occurring in the most central class of Pb-Pb collisions at $\sqrt{s_{NN}} = 2.76$ TeV. These phenomena seem to appear at the two levels: in individual events, where the production of identified hadrons in each collision can be describe within the chemical equilibrium SFOM but with the reduced content of the hadron gas, and in the whole sample, causing substantial differences among collisions belonging to the same most central class.

As last remarks, two more general, possible implications of the presented approach are pointed out. The rigorous treatment of the data ''points'' might shed some light on the excellent performance of the Tsallis distribution in the fitting of transverse momentum spectra observed at the RHIC and the LHC \cite{Azmi:2015xqa}. According to Ref.~\cite{Wilk:1999dr} the Tsallis distribution is the mean of the exponential (Boltzmann) distribution $\exp{(-\beta E)}$ weighted with a gamma distribution of $\beta$. In the view of this work, the Tsallis distribution (as a mean) is the more correct equivalent of the data than the Boltzmann distribution. Last, but not least, a great deal of data in high energy physics are averages, so in any theoretical modeling (of these data) one should be aware of possible misinterpretations when an average is compared with a prediction for a single event.

\section*{Acknowledgments}

The author thanks Tobias Fischer, Krzysztof Graczyk and Sebastian Szkoda for the help
in numerics. This work was done during a sabbatical at the home institution and was not supported by any financial grant. Most of calculations were carried out using resources provided by Wroclaw Centre for Networking and Supercomputing (http://wcss.pl), grant No. 268.


\end{document}